\documentstyle[12pt,twoside,fleqn,espcrc1,amsmath,epsfig,wrapfig]{article}
\begin{document}

\pagestyle{empty}

\begin{flushleft}
\large
{SAGA-HE-144-99
\hfill Feb. 10, 1999}  \\
\end{flushleft}
 
\vspace{1.8cm}
 
\begin{center}
 
\LARGE{{\bf Polarized proton-deuteron Drell-Yan}} \\
\vspace{0.2cm}

\LARGE{{\bf processes and parton distributions}} \\

\vspace{1.3cm}
 
\LARGE
{S. Hino and S. Kumano $^*$} \\
 
\vspace{0.4cm}
  
\LARGE
{Department of Physics}         \\
 
\LARGE
{Saga University}      \\
 
\LARGE
{Saga 840-8502, Japan} \\

\vspace{2.0cm}
 
\Large
{Talk at the KEK-Tanashi International Symposium} \\

\vspace{0.2cm}

{on PHYSICS OF HADRONS AND NUCLEI} \\

\vspace{0.2cm}

{Tokyo, Japan, Dec. 14 -- 17, 1998} \\

\vspace{0.05cm}

{(talk on Dec. 16, 1998) }  \\
 
\end{center}
 
\vspace{0.7cm}

\vfill
 
\noindent
{\rule{6.0cm}{0.1mm}} \\
 
\vspace{-0.3cm}
\normalsize
\noindent
{* Email: 97sm16@edu.cc.saga-u.ac.jp, 
kumanos@cc.saga-u.ac.jp} \\

\vspace{-0.44cm}
\noindent
{\ \ \ Information on their research is available 
 at http://www2.cc.saga-u.ac.jp/saga-u/riko}  \\

\vspace{-0.44cm}
\noindent
{\ \ \ /physics/quantum1/structure.html.} \\

\vspace{+0.1cm}
\hfill
{\large to be published in Nuclear Physics A}

\vfill\eject
\setcounter{page}{1}
\pagestyle{plain}


\title{Polarized proton-deuteron Drell-Yan processes and parton distributions}
\author{S. Hino and S. Kumano
\address{Department of Physics, Saga University, 
                 Saga 840-8502, Japan} 
\thanks
{http://www2.cc.saga-u.ac.jp/saga-u/riko/physics/quantum1/structure.html.
 S.K. was partly supported by the Grant-in-Aid for Scientific Research
from the Japanese Ministry of Education, Science, and Culture under
the contract number 10640277.}}
\maketitle
\begin{abstract}
We show in general that there are 108 structure functions in the
proton-deuteron Drell-Yan processes. However, there exist only 22 finite
ones after integrating the cross section over the virtual-photon
transverse momentum $\vec Q_T$ or after taking the limit $Q_T\rightarrow 0$.
There are 11 new structure functions in comparison with the ones
of the proton-proton reactions, and they are related to the tensor
structure of the deuteron. Parton-model analyses indicate an important
tensor structure function $V_T^{UQ_0}$, which can be measured by
a quadrupole spin asymmetry.
The Drell-Yan process has an advantage over lepton reactions
in finding tensor polarized antiquark distributions.
We hope that our studies will be realized in the next-generation
RHIC-Spin project and other ones.
\end{abstract}

\section{Introduction}

Spin structure of the proton has been investigated through polarized
deep inelastic lepton scattering. It is now interesting to investigate
other aspects of spin physics.
For example, the deuteron is a spin-one particle, so that
it has tensor structure which does not exist in the proton.
The deuteron target is often used in the deep inelastic scattering;
however, the purpose is to extract the ``neutron" structure functions
in the deuteron.  We had better shed light on the deuteron
spin structure itself rather than just use it for finding the neutron
information. There are some initial studies on the spin-one structure
functions. In the lepton scattering on the deuteron, there exist
new structure functions, $b_1$, $b_2$, $b_3$, and $b_4$.
However, the polarized proton-deuteron (pd) reaction had not been 
studied at all in connection with the deuteron structure functions. 
The major purpose of this work is to investigate the general formalism
of the polarized pd Drell-Yan processes \cite{our1}. In particular, we
list all the possible structure functions in the reactions. 
Then, the processes are also analyzed in a naive parton model \cite{our2}.
Another purpose is to facilitate future deuteron projects
such as the polarized deuteron reactions at RHIC.

The tensor structure function is a new field of high-energy spin physics.
Because it has not been measured experimentally, it is a unique opportunity
to test our present understanding of spin physics. Furthermore, it may
reveal an unexpected aspect of hadron structure. There is a possibility
that the polarized proton-deuteron reaction is realized in
the future experimental projects.

\section{Structure functions and spin asymmetries}

\begin{figure}[htb]
\begin{minipage}[t]{50mm}
   \begin{center}
       \epsfig{file=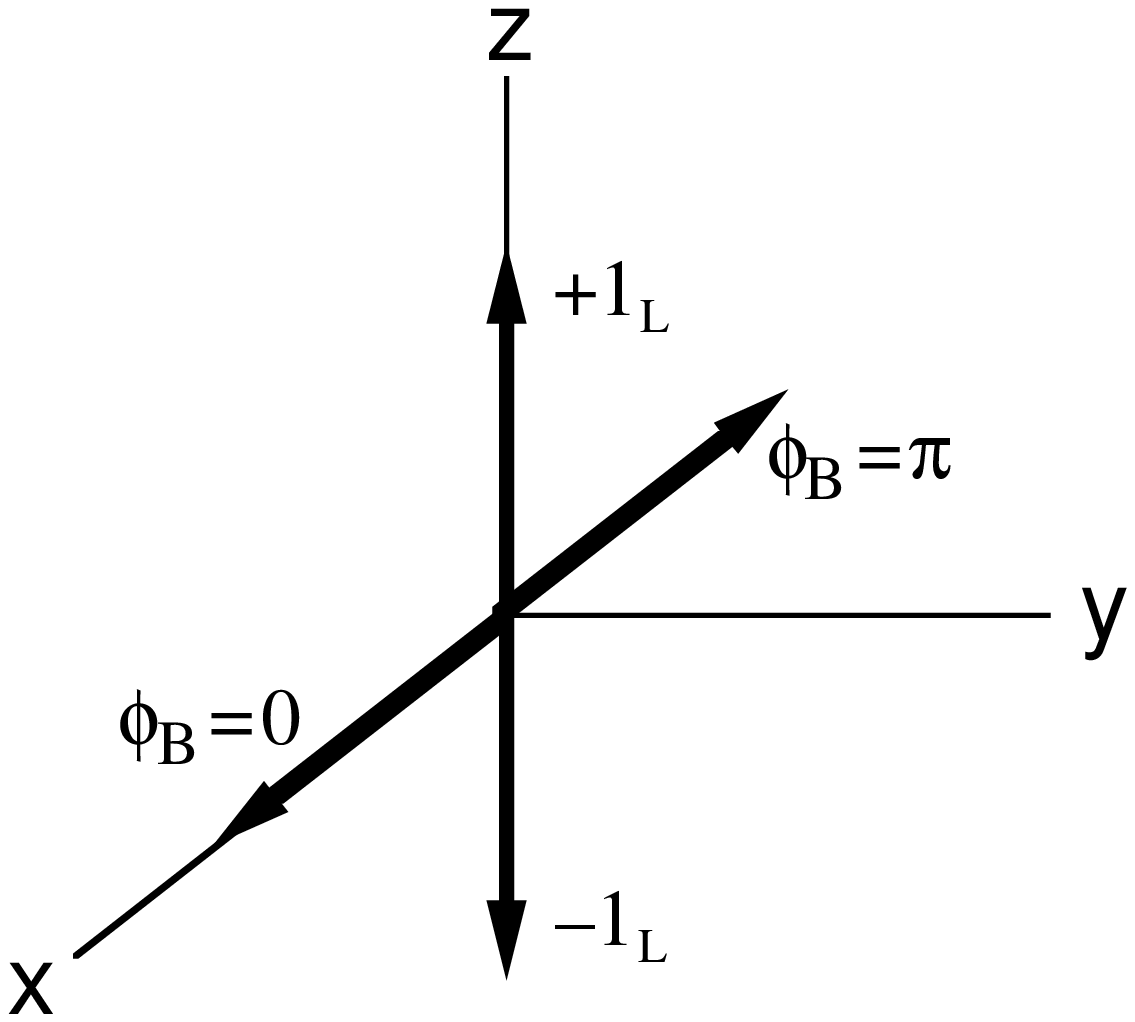,width=4.3cm}
   \end{center}
\vspace{-1.0cm}
\caption{Polarization $Q_0$.}
\label{fig:q0}
\end{minipage}
\hspace{\fill}
\begin{minipage}[t]{50mm}
   \begin{center}
       \epsfig{file=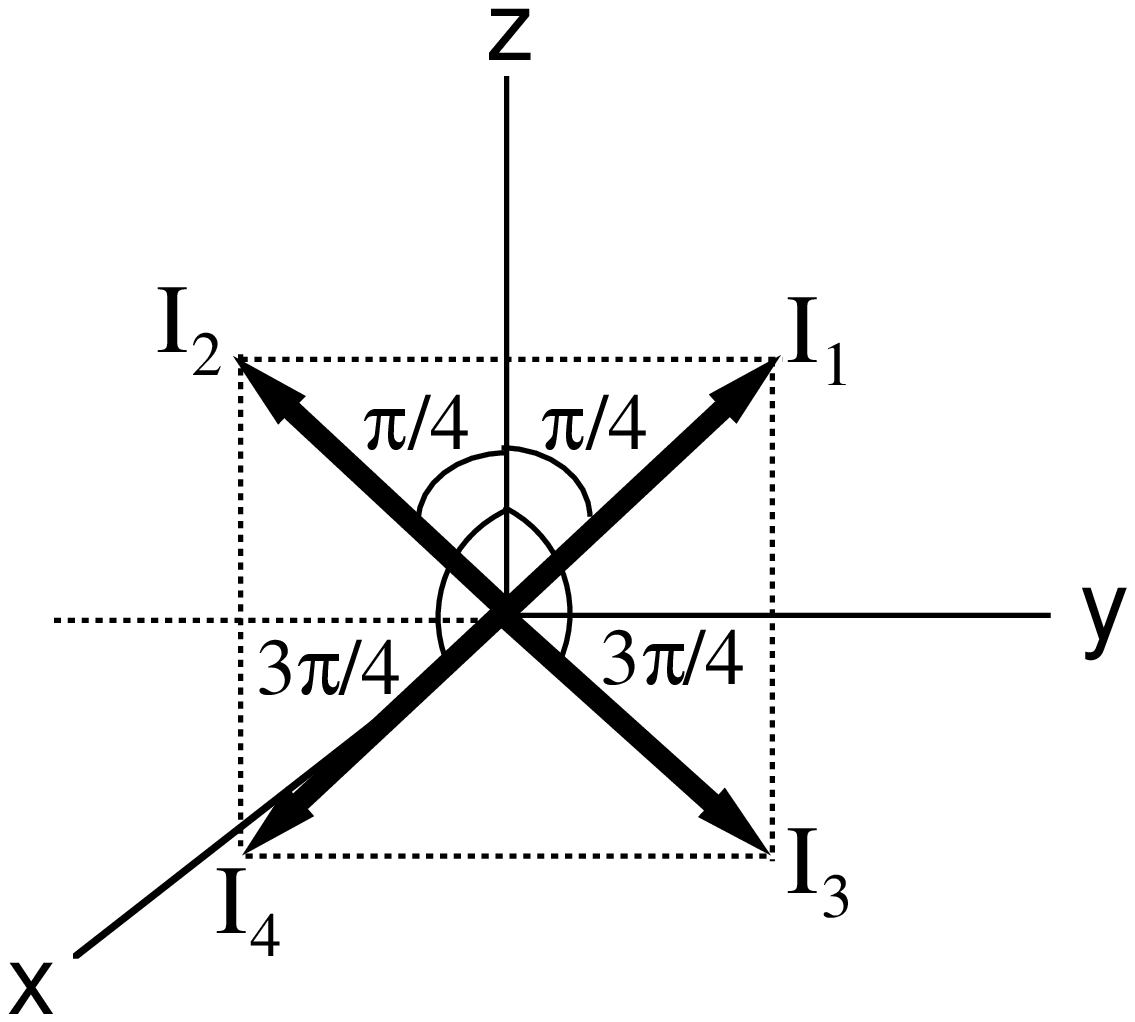,width=4.3cm}
   \end{center}
\vspace{-1.0cm}
\caption{Polarization $Q_1$.}
\label{fig:q1}
\end{minipage}
\hspace{\fill}
\begin{minipage}[t]{50mm}
   \begin{center}
       \epsfig{file=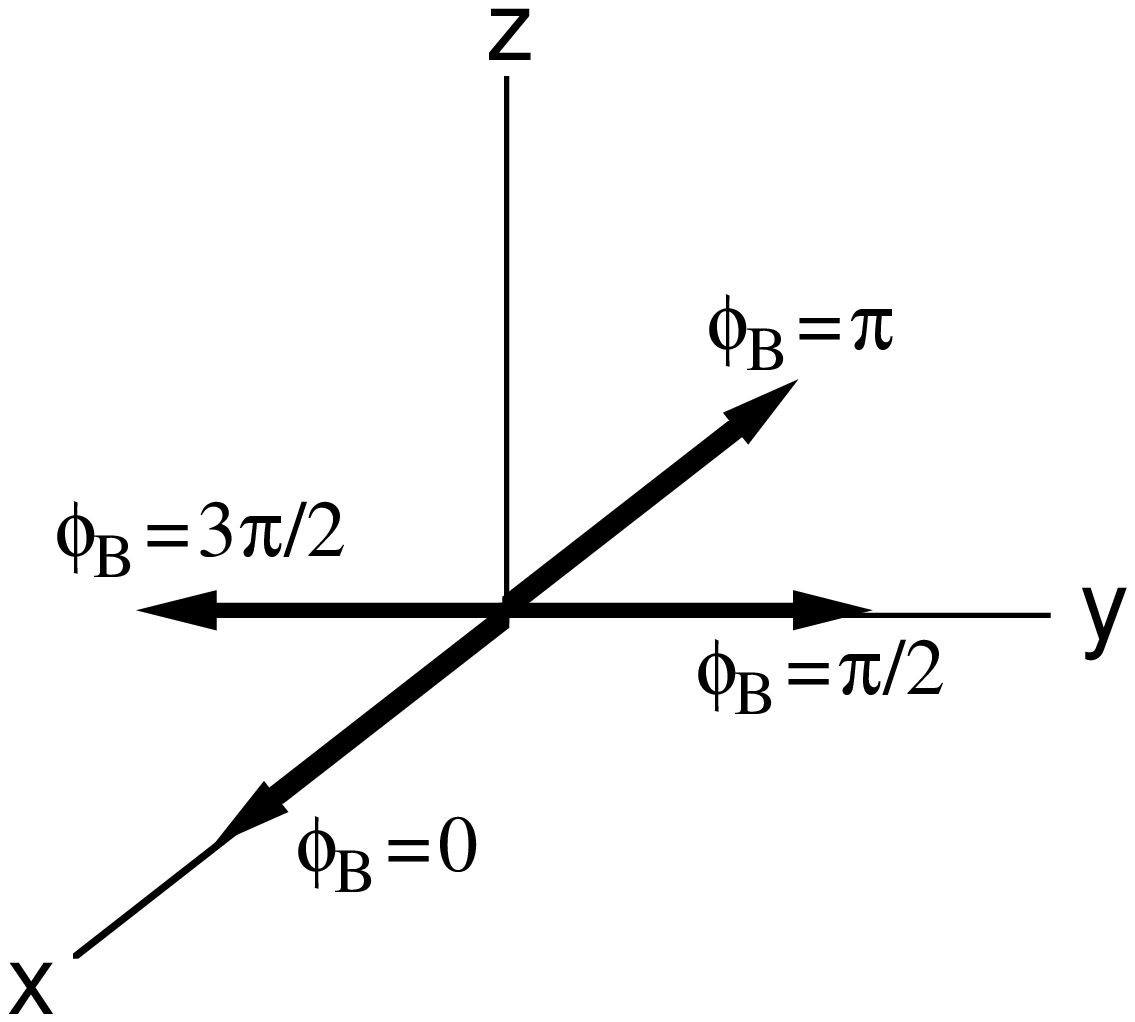,width=4.3cm}
   \end{center}
\vspace{-1.0cm}
\caption{Polarization $Q_2$.}
\label{fig:q2}
\end{minipage}
\end{figure}

Because the polarized high-energy pd Drell-Yan process had not been
studied, we should first discuss in general what kind of structure functions
are investigated \cite{our1}. Two independent formalisms are studied.
The first one is a spin-density formalism and the second is
a hadron-tensor formalism.
In the first method, we express structure functions by the helicity amplitudes
and Clebsch-Gordan coefficients. Then, imposing Hermiticity, parity
conservation, and time-reversal invariance, we find that 
108 structure functions exist. In comparison with the proton-proton (pp)
reactions, there are 60 new functions. If the cross section is integrated
over the lepton-pair transverse momentum $\vec Q_T$, there are only
22 ones and the others vanish. There are 11 new functions in addition
to the ones of the pp reactions, and they are associated with the tensor
structure of the deuteron. 
In the second method, possible Lorentz index expressions are
found for the  hadron tensor $W^{\mu\nu}$ by the combinations of
available Lorentz vectors and pseudovectors. The $Q_T\rightarrow 0$
case indicates that the following 22 structure functions also exist:
\begin{alignat}{11}
& W_{0,0} \, ,  
               & & V_{0,0}^{LL} ,    & & V_{0,0}^{TT} , 
               & & V_{0,0}^{U Q_0} , & & V_{0,0}^{TQ_1} , 
               & & W_{2,0} ,         & & V_{2,0}^{LL} ,
               & & V_{2,0}^{TT} ,    & & V_{2,0}^{U Q_0} ,
               & & V_{2,0}^{TQ_1} ,  & & U_{2,1}^{T U} ,  
\nonumber \\
& U_{2,1}^{TQ_0} , \ \ \     
               & & U_{2,1}^{U T} ,  \ \ 
               & & U_{2,1}^{LQ_1} , \ \ 
               & & U_{2,1}^{TQ_2} , \ \ 
               & & U_{2,1}^{TL} ,   \ \ 
               & & U_{2,1}^{LT}  ,  \ \ 
             & & U_{2,1}^{U Q_1}  , \ \ 
             & & U_{2,2}^{U Q_2} ,  \ \ 
             & & U_{2,2}^{TT}  ,    \ \ 
             & & U_{2,2}^{TQ_1} ,   \ \ 
             & & U_{2,2}^{LQ_2} .  
\label{eqn:sf}
\end{alignat}
The functions $W$, $V$, and $U$ are an unpolarized structure function,
a polarized one without the spin factors in the hadron tensor,
and a polarized one with the spin factor. 
The subscripts of these structure functions
indicate, for example, that $W_{L,M}$ is obtained by
$\int d\Omega \,  Y_{LM} \, d\sigma/(d^4Q \,d\Omega) \propto W_{L,M}$
in the unpolarized reaction. The superscripts indicate the polarization
states of $A$ and $B$: e.g. $U_{L,M}^{pol_A \, pol_B}$.
The superscripts $U$, $L$, and $T$ show unpolarized,
longitudinally polarized, and transversely polarized states.
The quadrupole polarizations $Q_0$, $Q_1$, and $Q_2$ are specific
in the reactions with a spin-1 hadron, and
they are associated with the spherical harmonics 
$Y_{20}$, $Y_{21}$, and $Y_{22}$ as shown
in Figs. \ref{fig:q0}, \ref{fig:q1}, and \ref{fig:q2}.
They are the polarizations in the $xz$, $yz$, and $xy$ planes.
The 11 structure functions with the superscripts $Q_0$, $Q_1$,
and $Q_2$ do not exist in the pp reactions. 

\vspace{-0.5cm}
\vspace{-0.6cm}

In the pp Drell-Yan processes, merely the following unpolarized,
longitudinal, and transverse combinations exist: $< \! \sigma \! >$,
$A_{LL}$, $A_{TT}$, $A_{LT}$, and $A_{T}$.
In addition, the quadrupole spin asymmetries exist in the pd reactions.
Our formalism indicates that the following fifteen quantities could
be investigated:
\begin{alignat}{8}
& < \! \sigma \! >, \ \ & & 
A_{LL}, \ \             & &
A_{TT}, \ \             & &
A_{LT}, \ \             & &
A_{TL}, \ \             & &
A_{UT}, \ \             & &
A_{TU}, \ \             & &
        \ \        \nonumber \\
& A_{UQ_0}, \ \         & &     
A_{TQ_0}, \ \           & &
A_{UQ_1}, \ \           & &
A_{LQ_1}, \ \           & &
A_{TQ_1}, \ \           & &
A_{UQ_2}, \ \           & &
A_{LQ_2}, \ \           & &
A_{TQ_2},
\label{eqn:asym}
\end{alignat}
where the subscript $U$ indicates the unpolarized state.
These asymmetries are expressed in terms of the structure functions
in Eq. (\ref{eqn:sf}). For example, $A_{UQ_0}$ is measured with
the unpolarized proton and the $Q_0$-type tensor polarized deuteron.
It is written by the structure functions $V_{0,0}^{UQ_0}$,
$V_{2,0}^{UQ_0}$, $W_{0,0}$, and $W_{2,0}$:
\begin{equation}
\! \! \! \! \! \! \! \! \! \! \! \! \! \! \! 
A_{UQ_0} = \frac{1}{2 < \! \sigma \! >} \,        
         \bigg [ \, \sigma(\bullet , 0_L)
            - \frac{ \sigma(\bullet , +1_L) 
                    +\sigma(\bullet , -1_L) }{2} \, \bigg ]   
         =  \frac{ 2 \, V_{0,0}^{UQ_0} 
                          + (\frac{1}{3}-cos^2 \theta ) \, 
                            V_{2,0}^{UQ_0} }
                 { 2 \, W_{0,0}
                     + (\frac{1}{3}-cos^2 \theta ) \,  W_{2,0}  }
\, ,
\label{eqn:a-uq0}
\end{equation}
where $\bullet$ indicates the unpolarized case.

\section{Parton-model analysis}

In the previous section, it is shown in Eq. (\ref{eqn:sf}) that there exist
the 22 structure functions. These functions are measured by the spin
asymmetries in Eq. (\ref{eqn:asym}).
Next, we try to understand the physics meaning of
these structure functions, particularly the new tensor polarized ones, 
in a naive parton model \cite{our2}. 

\begin{wrapfigure}{r}{0.42\textwidth}
   \vspace{-0.6cm}
   \begin{center}
       \epsfig{file=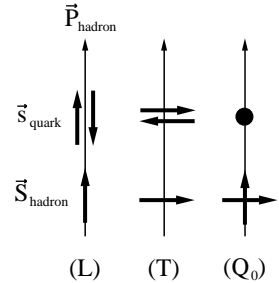,width=4.0cm}
   \end{center}
   \vspace{-0.5cm}
   \caption{Longitudinally polarized, transversity, 
              and tensor polarized distributions ($\bullet$=unpolarized).}
       \label{fig:ltq}
\end{wrapfigure}
The hadron tensor $W^{\mu\nu}$ is written by the quark and antiquark
correlation functions. They are expressed in terms of possible
combinations of Lorentz vectors and pseudovectors with the constrains
of Hermiticity, parity conservation, and time-reversal invariance.
We find that new tensor distributions are involved in
the correlation function $\Phi [\gamma^+]$. It is particularly
important that the leading-twist tensor distribution $\delta q$ is
found in our analysis, and it is expressed by the 
``unpolarized" quark distribution [${\mathcal P}(x)_\lambda$] 
in a spin-1 hadron with helicity $\lambda$ as
\begin{equation}
\delta q (x)= \frac{1}{2} \, \bigg[ \, {\mathcal P}(x)_{\lambda=0} 
          -  \frac{ {\mathcal P}(x)_{\lambda=+1} 
                   +{\mathcal P}(x)_{\lambda=-1} }{2} \, \bigg]
\ .
\label{eqn:tensor}
\end{equation}
It agrees with the definition in the lepton-deuteron scattering \cite{elfe}.
In the naive parton-model analysis, we find 19 structure functions.
However, there exist only four finite structure functions
by the $\vec Q_T$ integration and the others vanish. 
In the following, the $\vec Q_T$ integrated results are shown. 
First, the cross section is
\begin{align}
\frac{d \sigma}{dx_A \, dx_B \, d \Omega} = \frac{\alpha^2}{4 \, Q^2} \,
      \bigg [ \,  & (1 + \cos^2 \theta) \bigg\{ \, 
            \overline W_T 
            + \frac{1}{4} \lambda_A \lambda_B \, \overline V_T^{\, LL}
            + \frac{2}{3}  \left( 2 \, |\vec S_{BT}|^2 - \lambda_B^2 \right)
                               \, \overline V_T^{\, UQ_0} \, \bigg\}
\nonumber \\
&  +  \sin^2 \theta \, |\vec S_{AT}| \, |\vec S_{BT}|
             \cos(2\phi-\phi_A-\phi_B) \, \overline U_{2,2}^{\, TT}
        \, \bigg ]
\ ,
\label{eqn:cross-int}
\end{align}
where $\overline W = \int d^2 \vec Q_T \, W$ and the same for
$\overline V$ and $\overline U$.
The four finite structure functions 
are expressed by the parton distributions in the process
$q$(in p)+$\bar q$(in d)$\rightarrow \ell^+ + \ell^-$ as
\begin{alignat}{2}
\overline W_T     & = \frac{1}{3} \sum_a e_a^2 \, 
                 q_a (x_A) \, \bar q_a(x_B) \ , \ \ \ \ \ \ 
& 
\overline V_T^{\, LL} & = - \frac{4}{3} \sum_a e_a^2 \, 
          \Delta q_a (x_A) \, \Delta \bar q_a (x_B)
\ , \nonumber \\
\overline U_{2,2}^{\, TT} & = \frac{1}{3} \sum_a e_a^2 \, 
          \Delta_T q_a (x_A) \, \Delta_T \bar q_a (x_B)  \ , \ \ \ \ \ \ 
& 
\overline V_T^{\, UQ_0} & 
            = \frac{1}{3} \sum_a e_a^2 \, 
                          q_a (x_A) \, \delta \bar q_a (x_B)
\ .
\label{eqn:sf-int}
\end{alignat}
The unpolarized, longitudinally-polarized, transversity, and
tensor-polarized distributions are expressed by $q_a$, $\Delta q_a$,
$\Delta_T q_a$, and $\delta q_a$ with the quark flavor $a$,
respectively.
The tensor distribution can be investigated by the unpolarized-quadrupole
$Q_0$ asymmetry
\begin{equation}
A_{UQ_0} = \frac{\overline V_T^{\, UQ_0}}{\overline W_T}
         = \frac{\sum_a e_a^2 \, 
                  \left[ \, q_a(x_A) \, \delta \bar q_a(x_B)
                          + \bar q_a(x_A) \, \delta q_a(x_B) \, \right] }
                {\sum_a e_a^2 \, 
                  \left[ \, q_a(x_A) \, \bar q_a(x_B)
                          + \bar q_a(x_A) \, q_a(x_B) \, \right] }
\ ,
\end{equation}
where the contribution from the process
$\bar q$(in A)+$q$(in B)$\rightarrow \ell^+ + \ell^-$ is added.
This equation suggests that the tensor polarized distributions
are found by the Drell-Yan experiment if the unpolarized distributions
are well known in the proton. In the large $x_F$ region, the term
$\bar q_a(x_A) \, \delta q_a(x_B)$ is neglected so that the
antiquark tensor polarization $\delta \bar q$ could be extracted.
It could be also found in the lepton scattering by studying a deviation
from the $b_1$ sum rule \cite{b1sum}; however,
the Drell-Yan process has an advantage over the lepton reactions
in the sense that the antiquark distributions $\delta q_a (x)$ are found
rather easily.  
In addition, the antiquark flavor asymmetries $\Delta \bar u/\Delta \bar d$
and $\Delta_T \bar u/\Delta_T \bar d$ can be investigated by combining
the pd Drell-Yan data with the pp data in the same way as the unpolarized case
\cite{skpr} because the polarized pd Drell-Yan formalism is now completed 
in Refs. \cite{our1,our2}.

\section{Summary}

We have found that there exist 108 structure functions in the
pd Drell-Yan processes; however,
the number becomes 22 after integrating the cross section
over the transverse momentum $\vec Q_T$ or
after taking the limit $Q_T\rightarrow 0$. 
The new structure functions are associated with
the tensor structure of the deuteron.
The naive parton-model analyses indicate the important
structure function $V_T^{UQ_0}$, which is expressed by
the unpolarized distributions of the proton and the tensor polarized
ones of the deuteron. The pd Drell-Yan is particularly
useful for finding the tensor polarized antiquark distributions.
There are a variety of interesting topics on the polarized pd reactions.
We hope that our studies will be realized experimentally.



\end{document}